\documentclass[iop]{emulateapj}
\usepackage{txfonts}
\usepackage{hyperref}
\pdfoutput=1
\hypersetup{
  hyperfootnotes=true,    
  colorlinks=true,        
  linkcolor=black,        
  citecolor=black,        
  urlcolor=blue,          
  breaklinks=true,        
  pdffitwindow=false,     
  pdfstartview={FitH},    
  pdfpagemode=UseNone,    
  pdfauthor={Moshe Elitzur}
}

\graphicspath{{Figs/}}

\def\about  {\hbox{$\sim$}}
\def\E#1{\hbox{$10^{#1}$}}
\def\mic   {\hbox{$\mu$m}}
\def\deg    {\hbox{$^\circ$}}
\def\ga     {\hbox{$\gtrsim$}}
\def\la     {\hbox{$\lesssim$}}
\def\No    {\hbox{$N_0$}}
\def\CT    {\hbox{$C_{\rm T}$}}
\def\NH    {\hbox{$N_{\rm H}$}}
\def\cs    {\hbox{cm$^{-2}$}}
\def\Spitzer {\hbox{\em Spitzer}}
\def\INTEGRAL {\hbox{\em INTEGRAL}}

\shorttitle{AGN Unification}
\shortauthors{Moshe Elitzur}
\slugcomment{ApJ Letters, accepted for publication February 8, 2012}

\begin{document}

\title{On the Unification of Active Galactic Nuclei}

\author{Moshe Elitzur}

\affil{Department of Physics and Astronomy, University of Kentucky, Lexington,
KY 40506-0055; moshe@pa.uky.edu}

\begin{abstract}
The inevitable spread in properties of the toroidal obscuration of active
galactic nuclei (AGNs) invalidates the widespread notion that type 1 and 2 AGNs
are intrinsically the same objects, drawn randomly from the distribution of
torus covering factors. Instead, AGNs are drawn \emph{preferentially} from this
distribution; type 2 are more likely drawn from the distribution higher end,
type 1 from its lower end. Type 2 AGNs have a higher IR luminosity, lower
narrow-line luminosity and a higher fraction of Compton thick X-ray obscuration
than type 1. Meaningful studies of unification statistics cannot be conducted
without first determining the intrinsic distribution function of torus covering
factors.
\end{abstract}

\keywords{infrared: galaxies -- galaxies: active -- galaxies: nuclei --
galaxies: Seyfert -- quasars: general}

\section{Introduction}

\begin{figure*}
  \centering
 \includegraphics[width=0.9\hsize,clip]{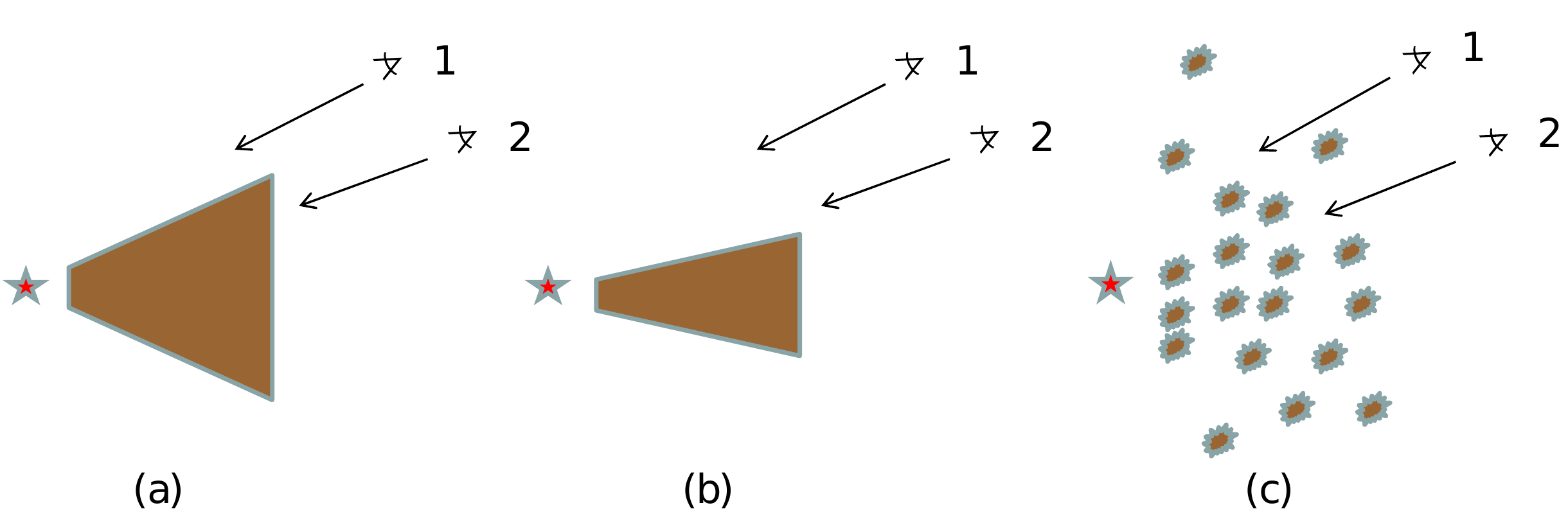}

\caption{AGN classification in unified schemes. ({\em a}) In a smooth-density
torus, everyone located inside the cone opening, such as observer 1, will see a
type 1 source; outside---a type 2. ({\em b}) Decreasing the torus covering
factor, the source becomes a type 1 AGN for more observers. ({\em c}) In a
clumpy, soft-edge torus, the probability for a direct view of the AGN decreases
away from the axis, but is always finite. }
 \label{fig:Cartoons}
\end{figure*}

The basic premise of the unification scheme is that all AGNs are fundamentally
the same: accreting supermassive black holes. The central engine is surrounded
by a dusty toroidal structure so that the observed diversity simply reflects
different viewing angles of an axisymmetric geometry \citep{Ski93, Urry95}. The
classification of AGNs into types 1 and 2 is based on the extent to which the
nuclear region is visible (figure \ref{fig:Cartoons}). Directions with clear
sight of the central engine and the broad-line region (BLR) yield type~1
sources. Those blocked by the torus result in type~2 objects, where the
existence of the hidden BLR is revealed only in polarized light. From basic
considerations, \cite{Krolik88} concluded that the torus likely consists of a
large number of individually very optically thick dusty clouds. Indeed, VLTI
interferometry of the Circinus AGN provide strong evidence for a clumpy or
filamentary dust structure \citep{Tristram07}.

In its most extreme form, dubbed the straw person model (SPM) by \cite{Ski93},
unification posits the viewing angle as the sole factor in determining AGN
classification; that is, (1) the classification of any AGN is determined
uniquely by its viewing angle and (2) the torus is identical for all AGNs of
the same luminosity.\footnote{Increasing luminosity tends to decrease the
covering factor as first noted by \cite{Lawrence91} in the ``receding torus"
model. For subsequent observations see \cite{Maiolino07a} and references
therein.} Clumpiness invalidates the first assumption because it turns the
difference between types 1 and 2 into the probability for direct view of the
AGN \citep{Elitzur07, Elitzur08, AGN2}. Since this probability is always
finite, type 1 sources can be detected from what are typically considered type
2 orientations, even through the torus equatorial plane. Conversely, if a cloud
happens to obscure the AGN from an observer, that object would be classified as
type 2 irrespective of the viewing angle (see figure \ref{fig:Cartoons}c). In
cases of such single cloud obscuration, on occasion the cloud may move out of
the line of sight, creating a clear path to the nucleus and a transition to
type 1 spectrum, as observed in a number of sources (see \citealt{Aret99} and
references therein).

The second assumption underlying SPM, about the torus sameness, obviously does
not hold. All AGNs cannot be expected to have the exact same torus; there must
be a spread in torus properties, even among AGNs with the same luminosity. Here
I discuss some immediate, fundamental implications of this inevitable spread.

\section{Realistic Unification}

In panel (a) of figure \ref{fig:Cartoons}, observer 1 will see the AGN as type
1, observer 2 as type 2. In panel (b) the AGN orientation is the same as in
panel (a), only its torus has a smaller covering factor. Now both observers see
a type 1 object even though their viewing angles have not changed. Evidently,
the torus covering factor \CT\ is as central to AGN classification as is the
viewing angle because an AGN with a larger covering factor has a higher
probability to be viewed as type 2 by a random observer. This is obvious also
for the more realistic clumpy torus, shown in panel (c) of figure
\ref{fig:Cartoons}. Therefore, in a sample of AGNs with a distribution of
covering factors, those with a larger \CT\ will have a higher probability to be
viewed as type-2 by a random observer, implying that
\begin{quote}
    {\em AGNs are drawn {\bf preferentially} from the distribution of covering
    factors; type-1 are more likely drawn from the distribution lower end, type-2
    from its higher end}.
\end{quote}
Contrary to the widespread notion that AGNs of types 1 and 2 are intrinsically
the same objects, fundamental differences between their average properties do
exist.

This is a more realistic formulation of the unification scheme than SPM, with
profound implications for AGN studies. Realistic unification immediately
explains the findings by \cite{RamosAlmeida09, RamosAlmeida11}, based on SED
modeling of AGN IR emission, that the dusty tori tend to have larger covering
factors in Seyfert 2 than in Seyfert 1. It also explains the seemingly puzzling
results of the recent study by \cite{Ricci11}, who analyzed in detail stacked
hard X-ray spectra (50--200 keV) of all $z < 0.2$ Seyfert galaxies detected
with \INTEGRAL, the hard X-ray/soft $\gamma$-ray mission. In agreement with the
basic tenets of unification, both Seyfert 1 and Seyfert 2 were found to have
the same average nuclear continuum emission, with a photon index of
$\Gamma$\,=\,1.8. But in apparent contradiction with unification, the
reflection component was significantly stronger for the average spectrum of
Compton thin Seyfert 2 than for Seyfert 1. Ricci et al find this discrepancy to
arise from a further sub-division among the Seyfert 2 AGNs. The ``lightly
obscured" ones (\NH\ $<$ \E{23} \cs) have the same reflection component as
Seyfert 1, $R\,\la\,0.4$, but those that are ``mildly obscured" ($\E{23}\,\cs
\le \NH < \E{24}\,\cs$) display a much stronger reflection with $R =
2.2^{+4.5}_{-1.1}$. While this finding contradicts simplistic forms of
unification, it is precisely the behavior expected from its realistic
formulation: Seyfert 1 and lightly obscured Seyfert 2 correspond to different
viewing angles of intrinsically similar AGNs, drawn from the low end of the
covering factor distribution, thus they conform, on average, to simplistic
unification. But in mildly obscured Seyfert 2 the absorber/reflector covers a
larger fraction of the X-ray source, producing stronger reflection that is not
seen in the average Seyfert 1 spectrum, where large covering factors are rare.
The large difference between the average reflection spectra of Seyfert 1 and 2
arises from significant differences in their torus covering factors.

\begin{figure}
  \centering
  \includegraphics[width=\hsize,clip]{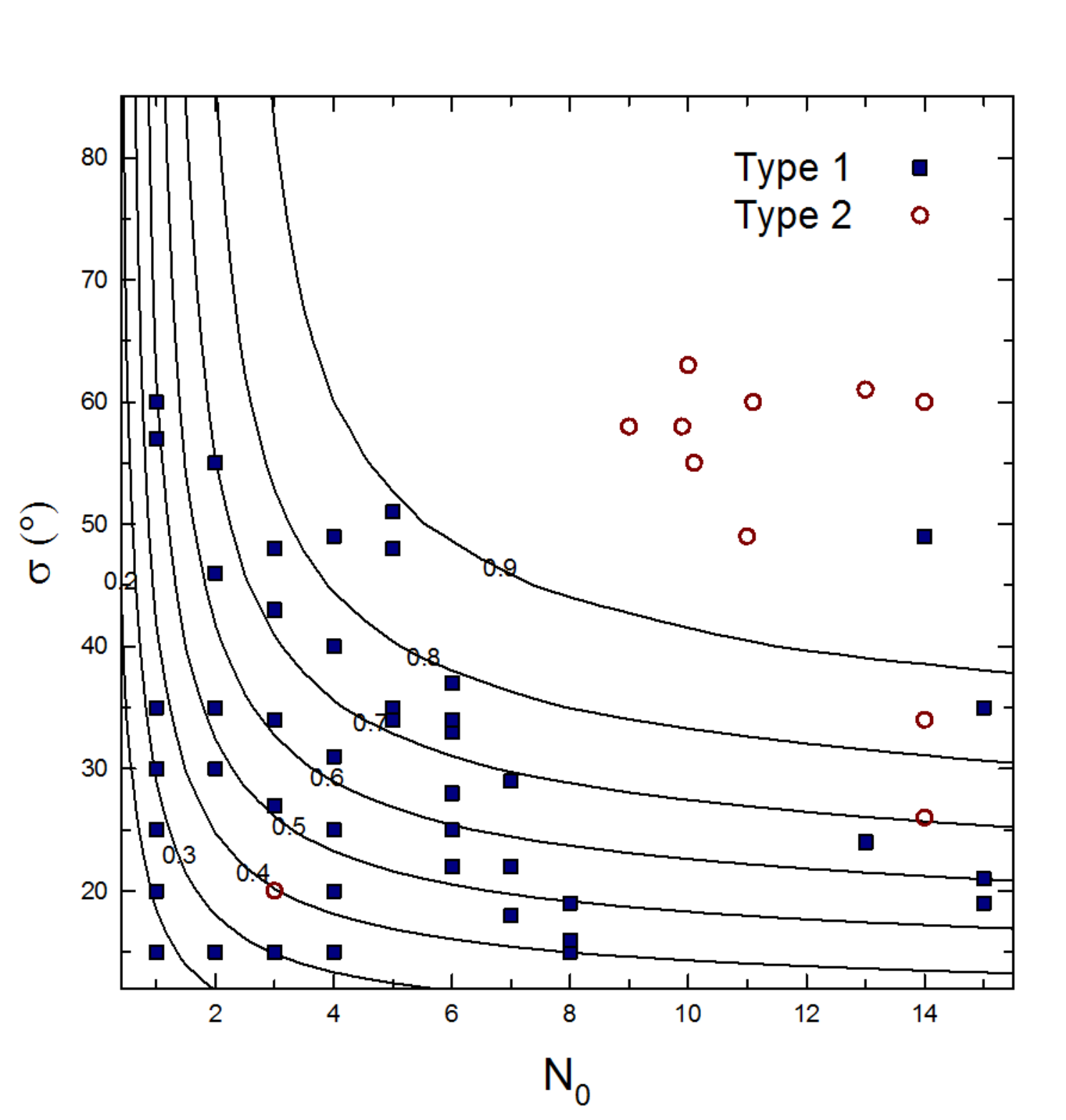}

\caption{Clumpy torus covering factors. Contour plots are for a toroidal
Gaussian distribution, where the number of clouds along viewing angle $i$ from
the axis is $N_0 e^{-(90 - i)^2/\sigma^2}$, with $N_0$ and $\sigma$ free
parameters. Each contour is the locus of $N_0$--$\sigma$ combinations that
produce the labeled covering factor. The data points are from clumpy torus
modeling of IR observations of AGNs reported in \cite{Mor09}, \cite{Nikutta09},
\cite{RamosAlmeida11}, \cite{Almudena11} and \cite{Deo11}.
\label{fig:Contours}}
\end{figure}


\section{Covering Factors}

The sometime loosely invoked concept of ``torus covering factor", \CT, can be
rigorously defined as the fraction of the sky at the AGN center covered by
obscuring material; it is the same as the fraction of randomly distributed
observers whose view to the center is blocked, and thus see a type 2 AGN
\citep{AGN1}. Denote by $N(i)$ the overall number of clouds encountered, on
average, along angle $i$ from the axis. Then the probability for direct viewing
of the AGN from that direction is $e^{-N(i)}$ and the torus covering factor is
$\CT = 1 - \int e^{-N(i)} d\cos i$. If \No\ is the average number of clouds
along radial equatorial rays then $N(i) = \No\Phi(i)$, where $\Phi(90\deg) =
1$. The cloud angular distribution function $\Phi$ can be conveniently
parametrized as Gaussian, $\Phi(i) = e^{-(90 - i)^2/\sigma^2}$, with $\sigma$
the distribution angular width \citep{Elitzur04, AGN2}. Fitting of IR
observations with clumpy torus models with Gaussian angular distributions has
been reported by a number of teams, and figure \ref{fig:Contours} shows the
results of these modeling efforts in the \No--$\sigma$ plane together with the
contour plots of \CT.\footnote{Earlier versions of this figure were presented
in \citeauthor{Elitzur09} (2009; accessible at
\url{www.mpe.mpg.de/events/pgn09/online_proceedings.html}) and
\cite{RamosAlmeida11}. The contour plots in both of these earlier figures are
afflicted by the computer bug reported in \cite{AGN2-Erratum}.} As expected
from realistic unification, and first noted by \cite{RamosAlmeida09}, type 1
and type 2 AGNs preferentially occupy different regions in the plane. The few
sources with a ``wrong" covering factor (large-\CT\ type 1, small-\CT\ type 2)
merely reflect the probabilistic nature of clumpy obscuration.  Although this
ad-hoc collection of AGNs, which were selected by different, unrelated
criteria, does not constitute a complete sample (only the \citealt{Mor09}
analysis of PG quasars involved a complete sample), it does illustrate the
point.

Since the covering factor measures the fraction of AGN luminosity captured by
the torus and converted to infrared, the AGN IR luminosity is $\CT L$, where
$L$ is its bolometric luminosity. Therefore {\em type 2 AGNs have intrinsically
higher IR luminosities than type 1}. Contrary to earlier expectations of strong
anisotropy at $\lambda\ \la\ 8\mic$, \Spitzer\ observations at this wavelength
regime show a great similarity between the IR fluxes of AGNs 1 and 2 when
normalized with either their X-ray fluxes \citep{Lutz04, Horst06} or optically
thin radio emission \citep{Buchanan06}. Part of this puzzle was solved by
clumpy torus calculations, which show much less anisotropy in IR emission than
the earlier smooth-density models \citep{AGN2}. Realistic unification explains
away the remainder.

While infrared arises from reprocessing of the AGN radiation captured by the
torus, narrow line emission is generated by the radiation that has {\em
escaped} the torus. The narrow line luminosity is proportional to $(1 - \CT)L$.
At the same bolometric luminosity, type 1 AGNs can be expected to have a higher
narrow line luminosity than type 2.


\section{Unification Statistics}

Implicitly or explicitly, all studies of AGN statistics assume that type 1 and
type 2 are intrinsically the same objects, drawn randomly from the distribution
of torus covering factors. Realistic unification invalidates this assumption,
implying that covering factors deduced from analysis of type 2 sources do not
necessarily apply to type 1 and vice versa. This may contribute to
discrepancies among studies of AGN statistics. Thus the fraction of Compton
thick AGNs (X-ray obscuring column \NH\,\ga\,\E{24}\,\cs) was determined to be
as high as 50\% in a pre-selected sample of Seyfert 2 \citep{Risaliti99,
Guainazzi05} but only \about~20\% in complete X-ray samples without
spectroscopic pre-selection \citep{Malizia09, Burlon11}.

\cite{Schmitt01} find that the type 2 fraction among Seyfert galaxies is \about
70\%, while \cite{Hao05b} find it to be only \about 50\%. This discrepancy can
be attributed, at least in part, to different selection methods. Both studies
based their selections on what they considered reliable isotropic properties:
Schmitt et al used 60\,\mic\ IR emission, Hao et al the narrow-line luminosity
of [O {\sc iii}] $\lambda5007$. However, as noted above, thanks to the larger
covering factor of their torus, Seyfert 2 convert a larger fraction of their
luminosity to IR, therefore IR-selection is biased in their favor. The opposite
afflicts the line selection criterion of Hao et al, which introduces preference
for smaller covering factors and type 1 AGNs.  Even though the emission itself
might be isotropic, in both cases it involves reprocessing of different
fractions of the intrinsic UV/optical continuum, and these fractions introduced
opposite biases in the two studies.

All previous findings involving unification statistics, including the synthesis
of the cosmic X-ray background (e.g., \citealt{Gilli07, Treister09}), therefore
need a critical reexamination and revision. One cannot draw statistical
inferences from AGN populations without folding in the intrinsic distribution
of torus covering factors, which is unknown. A reliable determination of this
distribution function requires IR modeling of an unbiased, complete sample of
AGNs. The only practical method to select AGNs with minimal classification bias
is through hard X-ray surveys; although such selection still misses the most
heavily obscured ($\NH > \E{25}\cs$) Compton thick AGNs \citep{Burlon11,
DeRosa12}, it is the least biased method of AGN selection. From the catalogue
of {\em INTEGRAL} observations in the 20--40 keV band, \cite{Malizia09}
extracted a complete sample of 88 AGNs, including type classification. Using
0.3--195 keV data from {\em Swift}--BAT observations and the {\em XMM-Newton}
archive, \cite{Burlon11} have compiled another complete sample that contains
199 type-classified AGNs. Identifying \Spitzer\ counterparts to AGNs in these
samples and fitting their IR observations with clumpy torus models, as was done
for the sources shown in figure \ref{fig:Contours}, is the only feasible
approach to determining the intrinsic distribution function of torus covering
factors. Searching through the \Spitzer\ archives shows that the Malizia et al
sample has 54 matches (L. Bassani and A. Malizia, private communication); a
similar search for the {\em Swift}--BAT sample can be expected to add
significantly to the number of blindly selected AGNs with \Spitzer\
counterparts. In a future publication we will report separately the results of
clumpy torus modeling of the IR emission for each source in this expanded list
to determine its torus covering factor from the fitted values of \No\ and
$\sigma$, and construct from the entire sample the probability distribution
function of \CT.

\section{Discussion}

Introducing SPM, \cite{Ski93} suggested that ``The true situation may be in
some sense half way between the SPM and the hypothesis that orientation doesn't
affect classification at all,'' ascribing classification to a single property
of the system --- orientation. However, quite apart from the indeterminism
introduced by clumpiness, AGN classification involves a fundamental plane
spanned by two independent axes --- orientation and covering factor. Simply
put, the true situation is not somewhere between SPM and no orientation
effects; it involves an additional, independent variable.

Because the covering factor is an intrinsic, observer-independent property, it
makes what has always been a difficult problem even more difficult. AGN
selection by an isotropic emission indicator is insufficient for reliable
statistics because most measured fluxes involve reprocessing of the intrinsic
UV/optical radiation and the fraction captured for reprocessing may differ. The
IR luminosity is $\CT L$ while the narrow line luminosity is proportional to
$(1 - \CT)L$, introducing biases in favor of, respectively, types 2 and 1. Past
studies of unification statistics cannot be fully trusted before they are
repeated taking into account the, as yet unknown, distribution of covering
factors.

\acknowledgements

I thank Loredana Bassani and Angela Malizia for help with their AGN sample.
Support by NASA and NSF is gratefully acknowledged.


\begin{thebibliography}{33}
\expandafter\ifx\csname natexlab\endcsname\relax\def\natexlab#1{#1}\fi

\bibitem[{{Alonso-Herrero} {et~al.}(2011){Alonso-Herrero}, {Ramos Almeida},
  {Mason}, {Asensio Ramos}, {Roche}, {Levenson}, {Elitzur}, {Packham},
  {Rodr{\'{\i}}guez Espinosa}, {Young}, {D{\'{\i}}az-Santos}, \&
  {P{\'e}rez-Garc{\'{\i}}a}}]{Almudena11}
{Alonso-Herrero}, A. {et~al.} 2011, \apj, 736, 82

\bibitem[{{Antonucci}(1993)}]{Ski93} {Antonucci}, R. 1993, \araa, 31, 473

\bibitem[{{Aretxaga} {et~al.}(1999){Aretxaga}, {Joguet}, {Kunth}, {Melnick}, \&
  {Terlevich}}]{Aret99}
{Aretxaga}, I., {Joguet}, B., {Kunth}, D., {Melnick}, J., \& {Terlevich}, R.~J.
  1999, \apjl, 519, L123

\bibitem[{{Buchanan} {et~al.}(2006){Buchanan}, {Gallimore}, {O'Dea}, {Baum},
  {Axon}, {Robinson}, {Elitzur}, \& {Elvis}}]{Buchanan06}
{Buchanan}, C.~L., {Gallimore}, J.~F., {O'Dea}, C.~P., {Baum}, S.~A., {Axon},
  D.~J., {Robinson}, A., {Elitzur}, M., \& {Elvis}, M. 2006, \aj, 132, 401

\bibitem[{{Burlon} {et~al.}(2011){Burlon}, {Ajello}, {Greiner}, {Comastri},
  {Merloni}, \& {Gehrels}}]{Burlon11}
{Burlon}, D., {Ajello}, M., {Greiner}, J., {Comastri}, A., {Merloni}, A., \&
  {Gehrels}, N. 2011, \apj, 728, 58

\bibitem[{{de Rosa} {et~al.}(2012){de Rosa}, {Panessa}, {Bassani}, {Bazzano},
  {Bird}, {Landi}, {Malizia}, {Molina}, \& {Ubertini}}]{DeRosa12}
{de Rosa}, A. {et~al.} 2012, \mnras, 420, 2087


\bibitem[{{Deo} {et~al.}(2011){Deo}, {Richards}, {Nikutta}, {Elitzur},
  {Gallagher}, {Ivezi{\'c}}, \& {Hines}}]{Deo11}
{Deo}, R.~P., {Richards}, G.~T., {Nikutta}, R., {Elitzur}, M., {Gallagher},
  S.~C., {Ivezi{\'c}}, {\v Z}., \& {Hines}, D. 2011, \apj, 729, 108

\bibitem[{{Elitzur}(2007)}]{Elitzur07} {Elitzur}, M. 2007, in ASP Conf. Ser.
    373: The Central Engine of Active
  Galactic Nuclei, ed. L.~C. {Ho} \& J.-M. {Wang}, 415--424

\bibitem[{{Elitzur}(2008)}]{Elitzur08} {Elitzur}, M. 2008, New Astronomy
    Review, 52, 274

\bibitem[{{Elitzur}(2009)}]{Elitzur09} {Elitzur}, M. 2009, in Physics of
    Galactic Nuclei, Ringberg Castle Workshop

\bibitem[{{Elitzur} {et~al.}(2004){Elitzur}, {Nenkova}, \&
  {Ivezi{\'c}}}]{Elitzur04}
{Elitzur}, M., {Nenkova}, M., \& {Ivezi{\'c}}, {\v Z}. 2004, in ASP Conf. Ser.
  320: The Neutral ISM in Starburst Galaxies, ed. S.~{Aalto},
  S.~{Huttemeister}, \& A.~{Pedlar}, 242--252

\bibitem[{{Gilli} {et~al.}(2007){Gilli}, {Comastri}, \& {Hasinger}}]{Gilli07}
    {Gilli}, R., {Comastri}, A., \& {Hasinger}, G. 2007, \aap, 463, 79

\bibitem[{{Guainazzi} {et~al.}(2005){Guainazzi}, {Matt}, \&
  {Perola}}]{Guainazzi05}
{Guainazzi}, M., {Matt}, G., \& {Perola}, G.~C. 2005, \aap, 444, 119

\bibitem[{{Hao} {et~al.}(2005){Hao}, {Strauss}, {Fan}, {Tremonti}, {Schlegel},
  {Heckman}, {Kauffmann}, {Blanton}, {Gunn}, {Hall}, {Ivezi{\'c}}, {Knapp},
  {Krolik}, {Lupton}, {Richards}, {Schneider}, {Strateva}, {Zakamska},
  {Brinkmann}, \& {Szokoly}}]{Hao05b}
{Hao}, L. {et~al.} 2005, \aj, 129, 1795

\bibitem[{{Horst} {et~al.}(2006){Horst}, {Smette}, {Gandhi}, \&
  {Duschl}}]{Horst06}
{Horst}, H., {Smette}, A., {Gandhi}, P., \& {Duschl}, W.~J. 2006, \aap, 457,
  L17

\bibitem[{{Krolik} \& {Begelman}(1988)}]{Krolik88} {Krolik}, J.~H., \&
    {Begelman}, M.~C. 1988, \apj, 329, 702

\bibitem[{{Lawrence}(1991)}]{Lawrence91} {Lawrence}, A. 1991, \mnras, 252, 586

\bibitem[{{Lutz} {et~al.}(2004){Lutz}, {Maiolino}, {Spoon}, \&
  {Moorwood}}]{Lutz04}
{Lutz}, D., {Maiolino}, R., {Spoon}, H.~W.~W., \& {Moorwood}, A.~F.~M. 2004,
  \aap, 418, 465

\bibitem[{{Maiolino} {et~al.}(2007){Maiolino}, {Shemmer}, {Imanishi}, {Netzer},
  {Oliva}, {Lutz}, \& {Sturm}}]{Maiolino07a}
{Maiolino}, R., {Shemmer}, O., {Imanishi}, M., {Netzer}, H., {Oliva}, E.,
  {Lutz}, D., \& {Sturm}, E. 2007, \aap, 468, 979

\bibitem[{{Malizia} {et~al.}(2009){Malizia}, {Stephen}, {Bassani}, {Bird},
  {Panessa}, \& {Ubertini}}]{Malizia09}
{Malizia}, A., {Stephen}, J.~B., {Bassani}, L., {Bird}, A.~J., {Panessa}, F.,
  \& {Ubertini}, P. 2009, \mnras, 399, 944

\bibitem[{{Mor} {et~al.}(2009){Mor}, {Netzer}, \& {Elitzur}}]{Mor09} {Mor}, R.,
    {Netzer}, H., \& {Elitzur}, M. 2009, \apj, 705, 298

\bibitem[{{Nenkova} {et~al.}(2008{\natexlab{a}}){Nenkova}, {Sirocky},
  {Ivezi{\'c}}, \& {Elitzur}}]{AGN1}
{Nenkova}, M., {Sirocky}, M.~M., {Ivezi{\'c}}, {\v Z}., \& {Elitzur}, M.
  2008{\natexlab{a}}, \apj, 685, 147

\bibitem[{{Nenkova} {et~al.}(2008{\natexlab{b}}){Nenkova}, {Sirocky},
  {Nikutta}, {Ivezi{\'c}}, \& {Elitzur}}]{AGN2}
{Nenkova}, M., {Sirocky}, M.~M., {Nikutta}, R., {Ivezi{\'c}}, {\v Z}., \&
  {Elitzur}, M. 2008{\natexlab{b}}, \apj, 685, 160

\bibitem[{{Nenkova} {et~al.}(2010){Nenkova}, {Sirocky}, {Nikutta},
  {Ivezi{\'c}}, \& {Elitzur}}]{AGN2-Erratum}
---. 2010, \apj, 723, 1827

\bibitem[{{Nikutta} {et~al.}(2009){Nikutta}, {Elitzur}, \& {Lacy}}]{Nikutta09}
    {Nikutta}, R., {Elitzur}, M., \& {Lacy}, M. 2009, \apj, 707, 1550

\bibitem[{{Ramos Almeida} {et~al.}(2011){Ramos Almeida}, {Levenson},
  {Alonso-Herrero}, {Asensio Ramos}, {Rodr{\'{\i}}guez Espinosa}, {P{\'e}rez
  Garc{\'{\i}}a}, {Packham}, {Mason}, {Radomski}, \&
  {D{\'{\i}}az-Santos}}]{RamosAlmeida11}
{Ramos Almeida}, C. {et~al.} 2011, \apj, 731, 92

\bibitem[{{Ramos Almeida} {et~al.}(2009){Ramos Almeida}, {Levenson},
  {Rodr{\'{\i}}guez Espinosa}, {Alonso-Herrero}, {Asensio Ramos}, {Radomski},
  {Packham}, {Fisher}, \& {Telesco}}]{RamosAlmeida09}
---. 2009, \apj, 702, 1127

\bibitem[{{Ricci} {et~al.}(2011){Ricci}, {Walter}, {Courvoisier}, \&
  {Paltani}}]{Ricci11}
{Ricci}, C., {Walter}, R., {Courvoisier}, T.~J.-L., \& {Paltani}, S. 2011,
  \aap, 532, A102+

\bibitem[{{Risaliti} {et~al.}(1999){Risaliti}, {Maiolino}, \&
  {Salvati}}]{Risaliti99}
{Risaliti}, G., {Maiolino}, R., \& {Salvati}, M. 1999, \apj, 522, 157

\bibitem[{{Schmitt} {et~al.}(2001){Schmitt}, {Antonucci}, {Ulvestad}, {Kinney},
  {Clarke}, \& {Pringle}}]{Schmitt01}
{Schmitt}, H.~R., {Antonucci}, R.~R.~J., {Ulvestad}, J.~S., {Kinney}, A.~L.,
  {Clarke}, C.~J., \& {Pringle}, J.~E. 2001, \apj, 555, 663

\bibitem[{{Treister} {et~al.}(2009){Treister}, {Urry}, \&
  {Virani}}]{Treister09}
{Treister}, E., {Urry}, C.~M., \& {Virani}, S. 2009, \apj, 696, 110

\bibitem[{{Tristram} {et~al.}(2007){Tristram}, {Meisenheimer}, {Jaffe},
  {Schartmann}, {Rix}, {Leinert}, {Morel}, {Wittkowski}, {R{\"o}ttgering},
  {Perrin}, {Lopez}, {Raban}, {Cotton}, {Graser}, {Paresce}, \&
  {Henning}}]{Tristram07}
{Tristram}, K.~R.~W. {et~al.} 2007, \aap, 474, 837

\bibitem[{{Urry} \& {Padovani}(1995)}]{Urry95} {Urry}, C.~M., \& {Padovani}, P.
    1995, \pasp, 107, 803

\end{thebibliography}

\end{document}